\documentclass[a4paper,12pt]{article}
\usepackage{amssymb,amsmath}
\usepackage{a4wide}
\usepackage{epsfig}

\newcommand{\RR}{{\mathbb{R}}}
\newcommand{\CC}{{\mathbb{C}}}

\newcommand{\ZZ}{{\mathbb{Z}}}
\newcommand{\pa}{\partial}
\newcommand{\ii}{{\mathrm{i}}}
\newcommand{\half}{\tfrac{1}{2}}

\newcommand{\diag}{\mathop{\mathrm{diag}}\nolimits}

\title{Symmetric Instantons and Discrete Hitchin Equations}
\author{R.\ S.\ Ward\footnote{email address: richard.ward@durham.ac.uk}
  \bigskip
  \\Department of Mathematical Sciences,
  \\Durham University, Durham DH1 3LE.}
\date{\today}

\begin{document}

\maketitle

\begin{abstract}
\noindent Self-dual Yang-Mills instantons on $\RR^4$ correspond to
algebraic ADHM data. The ADHM equations for $S^1$-symmetric instantons
give a one-dimensional integrable lattice
system, which may be viewed as an discretization of the Nahm equations.
In this note, we see that generalized ADHM data for $T^2$-symmetric
instantons gives an integrable two-dimensional lattice system, which may be
viewed as a discrete version of the Hitchin equations.
\end{abstract}

\section{Introduction}
The prototype for the idea of this paper is the well-known correspondence
between $S^1$-symmetric instantons (or hyperbolic BPS monopoles) and the
discrete Nahm equation. Recall that self-dual Yang-Mills instantons on
$\RR^4$ correspond to ADHM data \cite{ADHM78}, which consist of matrices
satisfying certain algebraic constraints. If we impose an $S^1$ symmetry on
the instantons, then the corresponding dimensional reduction gives hyperbolic
monopoles \cite{A84}, in other words BPS monopoles on hyperbolic 3-space
${\mathbb{H}^3}$. Such an $S^1$ action is classified by a positive
integer~$n$; and then the monopole mass, or equivalently the asymptotic
norm of the monopole Higgs field, is $n/2$.
For a given value of $n$, SU(2) hyperbolic monopoles of charge $k$ are the
same as SU(2) instantons of charge $nk$. With suitable scaling, the
$n\to\infty$ limit corresponds to the curvature of the hyperbolic space tending
to zero; in other words the hyperbolic monopole tends to a monopole on
$\RR^3$. Now BPS monopoles on $\RR^3$ correspond, via the Nahm
transform \cite{N83}, to solutions of the Nahm equation, which is a set
of ordinary differential equations on an interval of the real line. So one might
expect the $S^1$-symmetric ADHM constraints to be a discrete (lattice) version
of the Nahm equation, tending to it as $n\to\infty$; and this is exactly
what happens \cite{BA90}. This discrete Nahm equation, which is a
special case of the algebraic ADHM constraints, forms an integrable
one-dimensional lattice system \cite{W99, MS00}, 

The subject of the present paper is to extend this idea to the case
where there are two commuting circle symmetries rather than just one.
So the starting-point is $T^2$-symmetric Yang-Mills instantons on $\RR^4$, and
the corresponding $T^2$-symmetric ADHM data. Such a $T^2$-action is characterized
by a pair of positive integers $n_1$ and $n_2$. One special case which has
been known for a long time is where $n_1=1$ or $n_2=1$: this corresponds
to spherically-symmetric hyperbolic monopoles of unit charge \cite{N86, Ch86}.
The case of general $(n_1,n_2)$ was studied shortly afterwards
\cite{F90, FH90, A90}, mainly in the context of instantons invariant
under a finite cyclic group $\ZZ_p$. The relevant expressions for such
$T^2$-symmetric ADHM data are reviewed in section~2 below.
In section~3, however, we forget about $T^2$-symmetric instantons
as such, and focus on the constraint equations for the ADHM data.
It turns out that these may be interpreted as a two-dimensional lattice
version of the Hitchin equations \cite{L77, H87}, a gauge-theory system
on $\RR^2$ (or more generally on Riemann surfaces). This lattice system
is completely-integrable, in the sense of being the compatibility condition
for a Lax pair of lattice operators, and it tends to the Hitchin system of 
partial differential equations as $n_1,n_2\to\infty$.


\section{$T^2$-symmetric instantons}
The structure of the ADHM data for $T^2$-symmetric SU(2) instantons
was described in \cite{F90, FH90}, using the version of the ADHM
constraint equations due to Donaldson \cite{D84}. This section summarizes
the relevant aspects, in a form suitable for our purposes here.

The data for SU(2) instantons of charge $N$ consists \cite{D84} of four
complex matrices $(\alpha_1, \alpha_2, a, b)$, where the $\alpha_i$ are
$N\times N$, $a$ is $2\times N$, and $b$ is $N\times2$. They satisfy the
equation
\begin{equation}  \label{Don_eqn1}
  [\alpha_1, \alpha_2] +ba=0,
\end{equation}
and are required to be generic ({\sl ie.}\ to satisfy a maximal-rank condition).
The `gauge freedom' in this data is
\begin{equation}  \label{Don_gauge}
\alpha_i\mapsto p\alpha_i p^{-1}, \,\, a\mapsto qap^{-1},\,\, b\mapsto pbq^{-1},
\end{equation}
where $p\in$\,GL($N,\CC$) and $q\in$\,SU(2). The $(8N-3)$-dimensional
moduli space of instantons is the space of generic solutions of (\ref{Don_eqn1}),
factored out by (\ref{Don_gauge}). To convert $(\alpha_1, \alpha_2, a, b)$
into ADHM data, one also needs to solve
\begin{equation}  \label{Don_eqn2}
  [\alpha_1, \alpha_1^*] + [\alpha_2, \alpha_2^*]+bb^*-a^*a =0,
\end{equation}
which has an essentially-unique solution \cite{D84}. Here $b^*$ denotes
the complex-conjugate transpose of $b$.

Now the standard action of SO(4) on $\RR^4$ induces an action on
the space of instantons, and we are interested in instantons which are
invariant under the action of the maximal torus $T^2=S^1\times S^1$ in SO(4).
Such $T^2$-symmetric instantons are classified by a pair $(n_1,n_2)$
of positive integers, with $n_1n_2=N$. For each choice of $(n_1,n_2)$,
the solution space is 1-dimensional \cite{F90, FH90}.
From the instanton point of view this is because, given the imposed symmetry,
the only remaining free parameter is the instanton scale.
This in turn corresponds to an overall positive factor on the data
$(\alpha_1, \alpha_2, a, b)$.
Alternatively, thinking in terms of hyperbolic monopoles, we have a
rotationally-symmetric hyperbolic monopole with a fixed axis of symmetry,
and the free parameter is then the location of the monopole on its axis.
The hyperbolic monopole has mass $n_1/2$ and charge $n_2$. (From the
instanton point of view, this is the same solution as a hyperbolic monopole
of mass $n_2/2$ and charge $n_1$: the swap-map $n_1\leftrightarrow n_2$
has recently been used in the study of symmetric hyperbolic monopoles \cite{Co14}).

The corresponding $T^2$-symmetric ADHM-Donaldson data may be written
in the following form ({\it cf.}\ \cite{F90}). For a positive integer $n$, define
$n\times n$ matrices $E_j$ and $E_j^-$ by
\[
E_1=\diag(1,0,\ldots,0), \ldots, E_n=\diag(0,\ldots,0,1),
\]
\[
E_1^-=\diag_{-1}(1,0,\ldots,0), \ldots, E_{n-1}^-=\diag_{-1}(0,\ldots,0,1).
\]
Then set
\begin{equation}  \label{alpha12}
 \alpha_1=\sum_{j=1}^{n_2} \sum_{k=1}^{n_1-1} F_{j,k}\, E_j \otimes E_k^-,
  \quad
 \alpha_2=\sum_{j=1}^{n_2-1} \sum_{k=1}^{n_1} G_{j,k}\, E_j^- \otimes E_k,
\end{equation}
\begin{equation}  \label{ab}
    a=\left[\begin{matrix} 0 & \ldots & 0 & 0 \\ 
           0 & \ldots & 0 & a_0\end{matrix}\right], \quad
    b=\left[\begin{matrix} b_0 & 0 & \ldots & 0 \\ 
           0 & 0 & \ldots & 0 \end{matrix}\right]^t.
\end{equation}
Here $A\otimes B$ denotes the Kronecker product of an $n_2\times n_2$ matrix
$A$ and an $n_1\times n_1$ matrix $B$. The variables
$F_{j,k}$ (for $1\leq j\leq n_2$, $1\leq k\leq n_1-1$),
$G_{j,k}$ (for $1\leq j\leq n_2-1$, $1\leq k\leq n_1$),
$a_0$ and $b_0$ are all positive real numbers.
Then the equations (\ref{Don_eqn1}) and (\ref{Don_eqn2}) become,
respectively,
\begin{equation}  \label{FGeqn1}
 F_{j+1,k}\,G_{j,k}=G_{j,k+1}\,F_{j,k} \mbox{ for $1\leq j\leq n_2-1$, $1\leq k\leq n_1-1$}, 
\end{equation}
\begin{eqnarray}
& & {}  F_{j,k-1}\, F^*_{j,k-1} - F^*_{j,k}\, F_{j,k} 
  + G_{j-1,k}\, G^*_{j-1,k} - G^*_{j,k}\, G_{j,k} \nonumber \\
    & & {} 
- a_0^* a_0 \,\delta(j-n_2)\delta(k-n_1) + b_0 b_0^* \,\delta(j-1)\delta(k-1)=0.
  \label{FGeqn2}
\end{eqnarray}
In equation (\ref{FGeqn2}), the indices have the range
$1\leq j\leq n_2$, $1\leq k\leq n_1$, but undefined terms are to be omitted;
for example, if $k=1$ then the first term is omitted, since there is no variable
$F_{j,0}$.
The `stars' are unnecessary in (\ref{FGeqn2}), since the variables are real numbers;
but we retain them because the variables will become complex matrices in
the next section.

The system (\ref{FGeqn1}, \ref{FGeqn2}) consists of $2n_1n_2-n_1-n_2+1$
equations for $2n_1n_2-n_1-n_2+2$ variables, and has a one-parameter family
of solutions, as mentioned previously.
For example, in the case $n_1=n_2=2$ we get five equations for six variables,
and the solution is $F_{11}=F_{21}=G_{11}=G_{12}=\lambda$,
$a_0=b_0=\lambda\sqrt{2}$, with $\lambda>0$ arbitrary.
The case $n_1=2$, $n_2=3$ is mentioned as an example in \cite{F90}.
For $n_1=2$, $n_2=4$ the solution is
\[
F_{11}=F_{41}=G_{31}=G_{12}=\lambda\sqrt{2}, \quad
F_{21}=F_{31}=\lambda,
\]
\[
G_{11}=G_{32}=2\lambda, \quad
G_{21}=G_{22}=\lambda\sqrt{3}, \quad a_0=b_0=\lambda\sqrt{6}.
\]
Several other cases can be solved explicitly, but for general $(n_1,n_2)$
the solution is not known explicitly (and may not be expressible
in radicals). A numerical solution for the case $n_1=n_2=50$ is illustrated
in the next section.

Finally in this section, let us consider the limit $n_2=n\to\infty$, with $n_1=2$
fixed. One way of interpreting this is as an axially-symmetric hyperbolic
monopole of charge $n$ on a fixed hyperbolic space, letting
$n\to\infty$ to obtain a hyperbolic magnetic disc; such a limit was recently
described in detail in \cite{BHS15} for the case $n_1=1$. Alternatively,
we may think of an axially-symmetric
2-monopole on a hyperbolic space with curvature
$-1/n^2$: then the limit $n\to\infty$ should yield the Nahm data for the
axially-symmetric 2-monopole on $\RR^3$. To take this limit,
we can follow the same pattern as in \cite{BA90}. 
Regard the index $j$ as labelling a one-dimensional lattice with lattice
spacing $1/n$, and put $F_{j,1}=f(s)$, $G_{j,1}=n+g(s)$, $G_{j,2}=n+h(s)$.
Then the limit $n\to\infty$ leaves us with the differential equations
\[
f'=(h-g)f, \quad g'=-\half f^2 =-h'.
\]
The relevant solution of this is
\[
   h(s)=\frac{\pi}{4}\tan(\pi s/2)=-g(s), \quad f(s)=\frac{\pi}{2}\sec(\pi s/2),
\]
which corresponds to the Nahm data for an axially-symmetric
2-monopole on $\RR^3$.


\section{Discrete Hitchin equations}
In the equations (\ref{Don_eqn1}, \ref{Don_eqn2}), the vectors $a$ and $b$
play the role of boundary terms, and the interior terms only involve $\alpha_i$.
If we focus on the interior equations by setting $a=b=0$, then the
remaining equations are, in effect, the self-dual Yang-Mills
equations reduced to zero dimensions.
The idea now is to forget about instantons as such, and simply
to regard (\ref{FGeqn1}), and (\ref{FGeqn2}) with $a_0=b_0=0$, namely
\begin{equation}  \label{FGeqn3}
 F_{j+1,k}\,G_{j,k}=G_{j,k+1}\,F_{j,k}, \quad
  F_{j,k-1}\, F^*_{j,k-1} 
  + G_{j-1,k}\, G^*_{j-1,k} = F^*_{j,k}\, F_{j,k} + G^*_{j,k}\, G_{j,k}
\end{equation}
as a two-dimensional lattice system. The objects $F_{j,k}$ and $G_{j,k}$
no longer need to be real numbers: instead, we allow them to be complex
$p\times p$ matrices. So the order of the factors in each of the terms of
(\ref{FGeqn3}) becomes important.
The claim is that the resulting system is an
integrable lattice version of the U($p$) Hitchin equations on $\RR^2$.

In what follows, we shall take $n_1=n_2=n$ for simplicity,
but it is straightforward to relax this condition. The question
of what boundary conditions one might want to add to (\ref{FGeqn3})
is left open for the moment.

The Hitchin equations \cite{L77, H87} may be thought of as the
self-dual Yang-Mills equations reduced to $\RR^2$, obtained by factoring
out two translations in $\RR^4$, and the resulting system is as follows.
Let $(x,y)$ denote the usual $\RR^2$ coordinates, $(A_x,A_y)$ a gauge
potential, and $(\Phi_1,\Phi_2)$ a pair of Higgs fields. Take the gauge group to
be U($p$), so that $(A_x,A_y,\Phi_1,\Phi_2)$ are antihermitian $p\times p$
matrices. Then the Hitchin equations are
\begin{equation}  \label{Heqns}
  F=[\Phi_1,\Phi_2], \quad D_x\Phi_1=-D_y\Phi_2,\quad D_x\Phi_2=D_y\Phi_1,
\end{equation}
where $F=\pa_x A_y-\pa_y A_x+[A_x,A_y]$ is the gauge field, 
and $D_x\Phi_j=\pa_x\Phi_j+[A_x,\Phi_j]$ (similarly for $D_y\Phi_j$) are
the covariant derivatives of $\Phi_j$.

The lattice equations (\ref{FGeqn3}) are a discrete version of (\ref{Heqns})
in the following sense. Let us take the 
continuum limit by extending the method of \cite{BA90} and the previous
section. Namely, put $x=j/n$ and $y=k/n$, write
\begin{equation} \label{CtmLimit}
  G_{j,k}=n-A_x(x,y)+\ii\Phi_1(x,y), \quad F_{j,k}=n-A_y(x,y)+\ii\Phi_2(x,y),
\end{equation} 
and take the limit $n\to\infty$. The result is the Hitchin system (\ref{Heqns}).

The lattice system (\ref{FGeqn3}) is integrable simply by
virtue of being a special case of the ADHM constraints, but one can also see
directly that it arises from a lattice Lax pair. This is analogous to the Lax pair
for the discrete Nahm equations \cite{W99, MS00}.
Let $X$ be the operator which steps forward in the first index,
namely $X:F_{j,k}\mapsto F_{j+1,k}$; and similarly let $Y$ be the operator which
steps forward in the second index. Define a pair of operators on lattice $p$-vectors
by
\begin{equation} \label{LaxPair}
\eth_1 = G_{j,k}^* X +\zeta F_{j,k-1} Y^{-1}, \quad
\eth_2 = F_{j,k}^* Y -\zeta G_{j-1,k} X^{-1},
\end{equation} 
where $\zeta$ is a complex parameter. Then $[\eth_1,\eth_2]=0$ for all $\zeta$
if and only if the equations~(\ref{FGeqn3}) hold. In other words, the lattice
system~(\ref{FGeqn3}) is an integrable discretization of the Hitchin
equations~(\ref{Heqns}). Note that if we replace $\eth_j$ by the equivalent operators
$\eth_1-n(1+\zeta)$ and $\eth_2-n(1-\zeta)$, and take the limit $n\to\infty$
using the expressions~(\ref{CtmLimit}), then we get the usual Lax pair
$\{(D_x+\ii\Phi_1)-\zeta(D_y-\ii\Phi_2), (D_y+\ii\Phi_2)+\zeta(D_x-\ii\Phi_1)\}$
for the Hitchin equations~(\ref{Heqns}).

Since the Hitchin system has a U($p$) gauge invariance, one might expect the discrete
version to have a local gauge invariance on the lattice, and this is indeed the case.
Indeed, the equations (\ref{FGeqn3}) are invariant under
\begin{equation} \label{Gauge}
G_{j,k}\mapsto G'_{j,k}=\Lambda_{j+1,k} G_{j,k} \Lambda_{j,k}^{-1}, \quad
F_{j,k}\mapsto F'_{j,k}=\Lambda_{j,k+1} F_{j,k} \Lambda_{j,k}^{-1},
\end{equation} 
where $\Lambda$ is a U($p$)-valued function on the lattice.

\begin{figure}[htb]
\begin{center}
\includegraphics[scale=0.7]{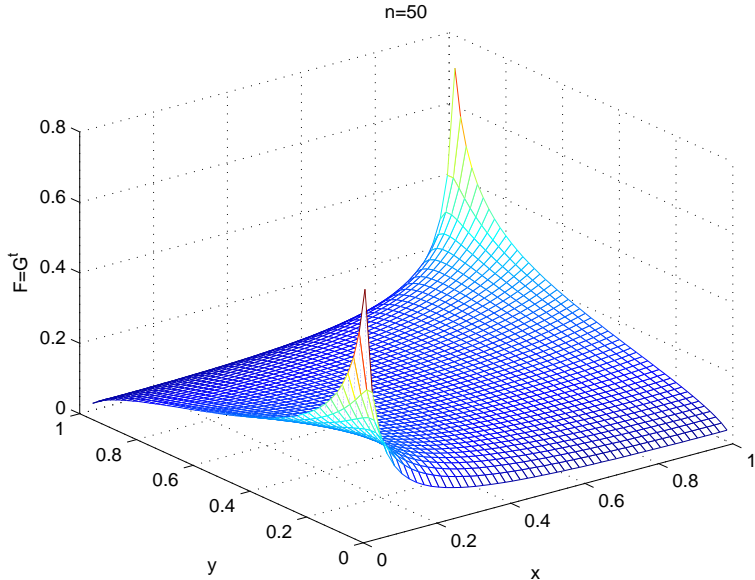}
\caption{Instanton data for $n_1=n_2=50$. \label{Fig1}}
\end{center}
\end{figure}

As an example, consider the simplest case $p=1$, so the gauge group is U(1).
The continuum system is then linear, and in fact reduces to the Cauchy-Riemann
equation $(\pa_x+\ii\pa_y)(\Phi_1-\ii\Phi_2)=0$. But the discrete system remains
nonlinear. One may choose a gauge such that $F_{j,k}$ and $G_{j,k}$ are
real-valued and positive, and the equations (\ref{FGeqn3}) then become
\begin{equation} \label{Eqns1_R}
\Delta_x^+ \log(F)=\Delta_y^+ \log(G), \,\, \Delta_y^- F^2 + \Delta_x^- G^2=0,
\end{equation} 
where
$\Delta_x^+$ denotes forward difference in the first index, 
$\Delta_x^-$ backward difference in the first index, and similarly
with $y$ referring to the second index. So from this point of view,
the system (\ref{Eqns1_R}) is an integrable nonlinear discretization
of the Cauchy-Riemann equations. One particular solution corresponds
to instanton data: for this one imposes boundary conditions as in
(\ref{FGeqn2}). Numerical solution of the equations
when $n_1=n_2=n$ indicates that this solution satisfies $G_{j,k}=F_{k,j}$
and $b_0=a_0$; and such a numerically-obtained solution is plotted in
Figure~\ref{Fig1} for the case $n=50$.

As yet, not much is known about the possible existence of explicit solutions
--- even in the simplest U(1) case --- except for small values of $n$. The simplest
one which is defined on the infinite lattice seems to be
\[
   F_{j,k}=\alpha\exp(\gamma j), \,\,  G_{j,k}=\beta\exp(\gamma k),
\]
where $\alpha$, $\beta$ and $\gamma$ are real constants. Whether
one can find more general families of explicit solutions is an open question.


\section{Concluding remarks.}
We have seen that the algebraic constraints on ADHM data for $T^2$-symmetric
instantons can be generalized
to give a set of lattice-gauge equations on a two-dimensional square lattice;
and this lattice system may be viewed as an integrable discrete version of
the Hitchin equations.

Since $S^1$-symmetric instantons may be interpreted as hyperbolic monopoles,
we may view $T^2$-symmetric instantons in that context as axially-symmetric
hyperbolic monopoles. But this treats the two circle actions differently, and
a more even-handed interpretation of $T^2$-symmetric instantons
is as a variant of the two-dimensional Hitchin system. Let us write the
$\RR^4$ coordinates $x_\mu$ as
$x_0+\ii x_3=r_1\exp(\ii\theta_1)$ and $x_1+\ii x_2=r_2\exp(\ii\theta_2)$,
and reduce by invariance in the $\theta_j$-directions. The gauge-potential
components $A_{\theta_j}$ then become Higgs fields $\Phi_j$. If we write
$A_j=A_{r_j}$, $\pa_j=\pa_{r_j}$ and $F=\pa_1A_2-\pa_2A_1+[A_1,A_2]$,
then the reduced self-dual Yang-Mills equations become
\begin{equation}  \label{Heqns_curved}
  F=(r_1r_2)^{-1}[\Phi_1,\Phi_2], \quad D_1\Phi_1=-(r_1/r_2)D_2\Phi_2.
   \quad D_1\Phi_2=(r_2/r_1)D_2\Phi_1,
\end{equation}
This is therefore an integrable variant of the standard Hitchin system
(\ref{Heqns}). Note that (\ref{Heqns_curved}) is invariant under
$r_j\mapsto\kappa r_j$, with $\Phi_j$ having conformal weight zero.
Solutions of (\ref{Heqns_curved}) which actually correspond
to instantons satisfy the boundary conditions
$|\Phi_j|\to\half n_j$ as $r_j\to0$ for $j=1,2$, in terms of the two integers $n_j$.
For each pair $(n_1,n_2)$, the equation (\ref{Heqns_curved}) has a
one-parameter of `instantonic' solutions, corresponding to the solutions of
(\ref{FGeqn1}, \ref{FGeqn2}). One open question is whether
less restrictive boundary conditions would allow more interesting
moduli spaces of solutions of (\ref{Heqns_curved}), no longer corresponding
to instantons.

Clearly many other questions remain open as well. For example,
it is likely that one could develop a more comprehensive treatment of
the complete-integrability of the discrete system, in particular through
understanding its spectral data, along the lines of what was done for the
discrete Nahm equations \cite{MS00}.

Another question is whether it is possible to impose boundary conditions
on the lattice system (\ref{FGeqn3}) which allow nice moduli spaces of solutions,
and/or a generalized Nahm transform. In the special case corresponding to
$T^2$-symmetric instantons, this Nahm transform is the ADHM transform;
but this case is rather special, with each solution space being just
one-dimensional. The Hitchin equations~(\ref{Heqns}) admit a rich collection of
moduli spaces of solutions, and whether any of this extends to the discrete
version remains open.

A final remark is that our starting-point above was the ADHM construction
for $S^1$- and $T^2$-symmetric instantons with gauge group SU(2). Recently
the $S^1$-symmetric case has been generalized to describe the
discrete Nahm equations corresponding to SU($N$) hyperbolic
monopoles \cite{Ch15}, and it might be interesting to see what
happens for $T^2$-symmetric instantons for larger gauge group.


\bigskip\noindent{\bf Acknowledgments.}
The author thanks the referee for comments which led to a substantial
revision of this paper, and acknowledges
support from the UK Particle Science and Technology
Facilities Council, through the Consolidated Grant No.\  ST/J000426/1.


\end{document}